\documentclass[a4paper,11pt]{article}
\usepackage[margin=2.5cm]{geometry}
\usepackage[utf8]{inputenc}
\usepackage{graphicx}
\usepackage{epsfig}
\usepackage{amsmath}
\usepackage{amssymb}
\usepackage{cite}
\usepackage{float}
\usepackage{subfigure}
\usepackage[font={footnotesize,it}]{caption}
\usepackage{authblk}
\numberwithin{equation}{section}
\usepackage{hyperref}
\hypersetup{
    colorlinks=true,
    linkcolor=blue,
    filecolor=red,      
    urlcolor=blue,
    citecolor=red
} 
\usepackage{epstopdf}
\DeclareGraphicsExtensions{.eps}
\usepackage{braket}
\usepackage{tikz}
\begin{document}

\begin{titlepage}

\title{ Holographic Entanglement Negativity for Adjacent Subsystems in $\mathrm{AdS_{d+1}/CFT_d}$}
\author[1,2,3]{Parul Jain\thanks{\noindent E-mail:~ parul.jain@ca.infn.it}}
\affil[1]{Dipartimento di Fisica, Universit\`a di Cagliari\\Cittadella Universitaria, 09042 Monserrato, Italy\smallskip}
\affil[2]{INFN, Sezione di Cagliari,Italy\bigskip}
\author[3]{Vinay Malvimat\thanks{\noindent E-mail:~ vinaymm@iitk.ac.in}} 
\author[3]{Sayid Mondal\thanks{\noindent E-mail:~  sayidphy@iitk.ac.in}}
\author[3]{Gautam Sengupta\thanks{\noindent E-mail:~  sengupta@iitk.ac.in}}
\affil[3]{Department of Physics\\

Indian Institute of Technology\\
 
Kanpur, 208016\\ 

India}

\maketitle

\abstract{
\noindent 
We establish our recently proposed holographic entanglement negativity conjecturefor mixed states of adjacent subsystems in conformal field theories with concrete higher dimensional examples. In this context we compute the holographic entanglement negativity for mixed states of adjacent subsystems in $d$-dimensional conformal field theories dual to bulk $AdS_{d+1}$ vacuum and $AdS_{d+1}$-Schwarzschild black holes.
These representative examples provide strong indication for the universality of our conjecture which affirms significant implications for diverse applications.
}
\end{titlepage}
\tableofcontents
\newpage
\section{Introduction}

Quantum entanglement has developed into an ubiquitous feature of modern fundamental physics in recent times connecting a spectrum of diverse areas from condensed matter physics to quantum gravity. In this regard entanglement entropy has evolved as the most significant and convenient measure to characterize the entanglement of a bipartite quantum system in a pure state. From quantum information theory this is defined as the von Neumann entropy of the reduced density matrix of the corresponding subsystem. 
For $(1+1)$-dimensional conformal field theories ($CFT_{1+1}$) the entanglement entropy may be computed 
through the {\it replica technique} developed by Calabrese et al in \cite{Calabrese:2004eu,Calabrese:2009qy}. 

In the context of the $AdS/CFT$ correspondence, Ryu and Takayanagi in \cite{Ryu:2006bv,Ryu:2006ef} proposed a prescription to compute the entanglement entropy of holographic $CFT$s. For a subsystem described by a spatial region $A$ on the boundary the entanglement entropy is given from this conjecture by the 
area of the co-dimension two bulk $AdS_{d+1}$ extremal surface anchored on the region $A$. 
In the recent past this has led to intense research activity into entanglement issues for diverse holographic $CFT$s both at zero and finite temperatures \cite{Takayanagi:2012kg,Nishioka:2009un,e12112244,blanco2013relative,Fischler:2012ca,Fischler:2012uv,Chaturvedi:2016kbk}. 

Although the entanglement entropy was crucial for the characterization of entanglement for bipartite systems in pure states, it was inadequate as an entanglement measure for mixed quantum states.  In a seminal work Vidal and Werner \cite{PhysRevA.65.032314} introduced a quantity termed {\it entanglement negativity} as a computable measure for the upper bound on the distillable entanglement in mixed states. The non convexity property of this entanglement measure was subsequently established in \cite{Plenio:2005cwa}. Interestingly, the authors in \cite{Calabrese:2012nk,Calabrese:2012ew,Calabrese:2014yza} computed this quantity in  ${CFT_{1+1}}$ employing a variant of the usual replica technique involving a certain four point function of the twist/anti-twist fields. This technique has been extensively employed to compute the entanglement negativity of various mixed state configurations in $CFT_{1+1}$\cite{Calabrese:2013mi,1367-2630-16-12-123020,PhysRevB.90.064401,Wen:2015qwa,PhysRevA.88.042319,1742-5468-2014-12-P12017}.

Naturally, it was critical to establish a holographic description for the entanglement negativity of boundary $CFT$s in terms of the bulk dual geometry in the $AdS/CFT$ scenario. In spite of interesting insights in \cite{Hubeny:2007xt,Rangamani:2014ywa}, a clear holographic prescription for the entanglement negativity of $CFT$s remained an unresolved issue. Two of the present authors (VM and GS) in the articles \cite{Chaturvedi:2016rcn,Chaturvedi:2016opa,Chaturvedi:2016rft} {\it (CMS)} proposed a holographic conjecture for the entanglement negativity of such boundary $CFT_d$s which exactly reproduced the $CFT_{1+1}$ \cite{Calabrese:2014yza} results in the large central charge limit.   

It is important to emphasize that the {\it CMS} conjecture mentioned above refers to the entanglement negativity of a single subsystem within an infinite system described by the boundary $CFT_d$. In the articles \cite{Calabrese:2012ew,Calabrese:2014yza}, the authors computed the entanglement negativity
of a mixed state characterized by two finite intervals $A_1$ and $A_2$ in a $CFT_{1+1}$ both at zero and finite temperatures. In a recent communication \cite{Jain:2017aqk}, the present authors  established an independent holographic conjecture for the entanglement negativity between the 
two intervals mentioned above in the context of $AdS_3/CFT_2$. It was shown there that the corresponding entanglement negativity was characterized by a certain algebraic sum of the geodesic lengths in the bulk $AdS_3$ space time anchored on the two adjacent intervals, which reduced to the holographic mutual information. Remarkably the holographic entanglement negativity computed from the above prescription exactly reproduced the $CFT_{1+1}$ results both for zero and finite temperatures in the large central charge limit \cite {Calabrese:2012ew,Kulaxizi:2014nma}. The holographic conjecture for the entanglement negativity  \cite {Jain:2017aqk} alluded above allowed a direct generalization to the $AdS_{d+1}/CFT_d$ scenario. In this case the entanglement negativity could be characterized in terms of an algebraic sum of the areas of bulk co-dimension two extremal surfaces anchored on the respective subsystems in the boundary $CFT_d$. As earlier this reduces to the holographic mutual information between the subsystems.

In this article we provide the first non trivial higher dimensional examples in the context of the $AdS_{d+1}/CFT_d$ correspondence to establish the efficacy of our conjecture. To this end we consider the mixed state of two adjacent subsystems $ A_1$ and $A_2$ characterized by rectangular strip geometries and compute the corresponding holographic entanglement negativity in $CFT_d$s at both zero and finite temperatures. For zero temperature the bulk configuration is described by the $AdS_{d+1}$ vacuum 
whereas the finite temperature scenario is described by the $AdS_{d+1}$-Schwarzschild black hole. For the finite temperature case the computation of the holographic entanglement negativity requires both a low and a high temperature approximations for the areas of the corresponding bulk extremal surfaces. At low temperatures the leading contribution arises from the $AdS_{d+1}$ vacuum corrected by sub leading thermal contributions. Interestingly for the high temperature case on the other hand the thermal contribution are precisely subtracted out. Hence at the leading order the entanglement negativity at high temperature is characterized by the area of the entangling surface on the boundary.

This article is organized as follows, in section $2$ we briefly review 
the computation of the holographic entanglement negativity of two adjacent intervals in the $AdS_3/CFT_2$ scenario described in \cite {Jain:2017aqk}. In section $3$ we establish the corresponding holographic conjecture for the entanglement negativity of two adjacent subsystems in the context of the $AdS_{d+1}/CFT_d$ correspondence. In the subsequent section $4$ we employ our conjecture to compute the holographic entanglement negativity for two adjacent subsystems of rectangular strip geometries at zero temperature from the bulk $AdS_{d+1}$ vacuum. In section $5$ we describe the corresponding computation for the finite temperature scenario from a bulk $AdS_{d+1}$-Schwarzschild black hole. In the final section $6$ we summarize our results and present our conclusions and future open issues.
\section{Entanglement negativity in $\mathrm{CFT_{1+1}}$}
\begin{figure}[H]\label{Figure1}
\begin{center}
\begin{tikzpicture}
\draw(-3,0)--(3,0); 
\draw(-2,0.3)--(-2,-0.3);
\draw(-0.7,0.3)--(-0.7,-0.3);
\draw(0.7,0.3)--(0.7,-0.3);
\draw(2,0.3)--(2,-0.3);
\node()at (-2.3,0.5){B};
\node()at (2.3,0.5){B};
\node()at (0,0.5){B};
\node()at (-1.3,0.5){$A_1$};
\node()at (-1.3,-0.3){$l_1$};
\node()at (-2,-0.5){$u_1$};
\node()at (-0.7,-0.5){$v_1$};
\node()at (1.3,0.5){$A_2$};
\node()at (1.3,-0.3){$l_2$};
\node()at (0.7,-0.5){$u_2$};
\node()at (2,-0.5){$v_2$};
\node()at (0,-1.5){};
\end{tikzpicture}
\caption{Schematic of two disjoint intervals $A_1$ and $A_2$ in a $(1+1)$-dimensional boundary $\mathrm{CFT}$.}\label{fig1}
\end{center}
\end{figure}
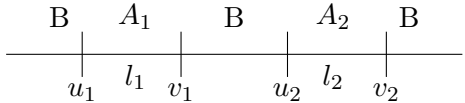

In this section, we briefly recapitulate the essential elements for the {\it entanglement negativity} of mixed states in a $CFT_{1+1}$~ \cite{Calabrese:2012nk,Calabrese:2012ew}. To this end we consider a tripartition in the  $CFT_{1+1}$ described by the spatial intervals $A_1$, $A_2$ and $B$ with $A=A_1 \cup A_2=[u_1,v_1]\cup[u_2,v_2]$, and $B=A^c$ represents the rest of the system\footnote{Note that the definition of entanglement negativity requires the concept of purification which involves embedding the given bipartite system ($A_1\cup A_2=A$) in a mixed state, inside a larger system $B$ such that the full system $A_1\cup A_2\cup B$ is in a pure quantum state. The larger system $B$ is then traced out to obtain the required mixed state $\rho_{A}$ of the bipartite quantum system.  }. The reduced density matrix of the subsystem $A$ is defined as $\rho_{A}=\mathrm{Tr}_{B} ~\rho $   and  $\rho_{A}^{T_2}$ is the partial transpose of the reduced density matrix with respect to the interval $A_2$. The entanglement negativity~ $\mathcal{E}$ is defined as the logarithm of the trace norm of the partially transposed reduced density matrix \cite{PhysRevA.65.032314}, which is expressed as
\begin{equation}\label{EN}
\mathcal{E}  = \ln \mathrm{Tr}|\rho_A^{T_2}|.
\end{equation}
The entanglement negativity may now be obtained through a replica technique as discussed in \cite{Calabrese:2012nk,Calabrese:2012ew} to determine $ \mathrm{Tr}~(\rho_A^{T_2})^{n_e}$ and the replica limit is given as the analytic continuation
of $n_e$ through even sequences to $n_e\to1$. This leads to the following expression for the entanglement negativity
\begin{equation}\label{ENRT}
\mathcal{E} = \lim_{n_e \rightarrow 1 } \ln \mathrm{Tr}(\rho_A^{T_2})^{n_e}.
\end{equation}
For the mixed state described by the two intervals as shown in Fig. (\ref{fig1}), the quantity $\mathrm{Tr}(\rho_A^{T_2})^{n_e}$ is given by a four point function of the twist operators on the complex plane from the replica technique described in \cite {Calabrese:2012nk,Calabrese:2012ew}, as follows
\begin{equation}\label{four point}
\mathrm{Tr}(\rho_A^{T_2})^{n_e} = 
\langle\mathcal{T}_{n_e}(u_1)\overline{\mathcal{T}}_{n_e}(v_1)\overline{\mathcal{T}}_{n_e}(u_2)\mathcal{T}_{n_e}(v_2)\rangle_{\mathbb{C}}.
\end{equation}
\subsection{Entanglement negativity for two adjacent intervals in vacuum}
\begin{figure}[H]
\begin{center}
\begin{tikzpicture}
\draw(-3,0)--(3,0); 
\draw(-1.5,0.3)--(-1.5,-0.3);
\draw(0,0.3)--(0,-0.3);
\draw(1.5,0.3)--(1.5,-0.3);
\node()at (-2.3,0.5){B};
\node()at (2.3,0.5){B};
\node()at (-1.5,-0.5){$u_1$};
\node()at (0,-0.5){$u_2$};
\node()at (1.5,-0.5){$v_2$};
\node()at (-0.8,0.5){$A_1$};
\node()at (-0.8,-0.5){$l_1$};
\node()at (0.8,0.5){$A_2$};
\node()at (0.8,-0.5){$l_2$};
\node()at (0,-1.5){};
\end{tikzpicture}
\caption{Schematic of two adjacent intervals $A_1$ and $A_2$ in a $(1+1)$-dimensional $\mathrm{CFT}$.}\label{fig2}
\end{center}
\end{figure}
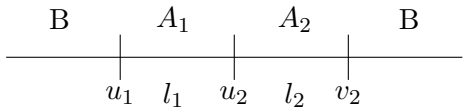

We first review the computation of the entanglement negativity for the mixed state of  two adjacent intervals in a $CFT_{1+1}$ at zero temperature \cite{Calabrese:2012nk,Calabrese:2012ew} and the corresponding holographic description in \cite {Jain:2017aqk}. The related configuration may now be obtained by setting $v_1\rightarrow u_2$ with $u_2=0,~u_1=-l_1$ and $v_2=l_2$ as shown in Fig. (\ref{fig2}) described below.
The quantity $\mathrm{Tr}(\rho_A^{T_2})^{n_e}$ in the eq. (\ref{four point}) for the two adjacent intervals is now described by a three point function of the twist operators as follows
\begin{equation}\label{trace 3-point}
\mathrm{Tr}(\rho_A^{T_2})^{n_e} = 
\langle\mathcal{T}_{n_e}(-l_1)\overline{\mathcal{T}}^2_{n_e}(0)\mathcal{T}_{n_e}(l_2)\rangle.
\end{equation}
The replica limit $n_e \to 1$ on eq. (\ref{trace 3-point}) now leads to the following expression for the entanglement negativity
\begin{equation}\label{cft negativity}
\mathcal{E} = \frac{c}{4} \ln \Big(\frac{l_1l_2}{(l_1+l_2)a}\Big)+ \mathrm{const},
\end{equation}
where $a$ is the UV cutoff for the $CFT$. The `const' term in the above expression may be neglected in the large central charge limit ( see discussion below) \cite{Kulaxizi:2014nma,Jain:2017aqk}. 

In \cite {Jain:2017aqk} the present authors demonstrated that the universal part of the three point function in eq.(\ref{trace 3-point}) is dominant in the large central charge limit and factorizes into two point correlation functions. Employing the geodesic approximation for these two point functions from the standard $AdS/CFT$ dictionary  then leads us to a holographic conjecture for the entanglement negativity 
of the configuration described above. In the context of the $AdS_3/CFT_2$ scenario, the holographic conjecture may be expressed as follows
\begin{equation}\label{HEN CONJ1}
\mathcal{E} = \frac{3}{16G^{(3)}_N}(\mathcal{L}_{A_1}+\mathcal{L}_{A_2}-\mathcal{L}_{A_1\cup A_2}).  
\end{equation}
Here $G^{(3)}_N$ is the $(2+1)$-dimensional Newton constant and $\mathcal{L}_{A_i}$ is the geodesic length anchored on the interval $A_i$. Using the Ryu-Takayanagi conjecture \cite{Ryu:2006ef,Ryu:2006bv} the eq. (\ref{HEN CONJ1}) reduces to the following
\begin{equation}\label{mutual inf}
\mathcal{E} =  \frac{3}{4}(S_{A_1}+S_{A_2}-S_{A_1\cup A_2})=\frac{3}{4}[{\cal I}(A_1,A_2)],
\end{equation}
which is precisely the mutual information between the subsystems described by the intervals $A_1$ and $A_2$. Note that the entanglement negativity is a measure of the upper bound on the distillable entanglement of the bipartite system whereas the mutual information is the upper bound on the total correlations between the subsystems. Therefore they are measures of distinct quantities in quantum information theory. However the universal parts of both are dominant in the large central charge limit and admit a holographic description which match exactly for this particular mixed state configuration \footnote{Note that recently this matching between the universal parts of  entanglement negativity and mutual information for the adjacent interval case  has also been observed in both local and global quench problems in a $CFT_{1+1}$  \cite{Coser:2014gsa,Wen:2015qwa}. }. Note that the full entanglement negativity and the mutual information involve non universal terms which are sub leading in 
$\frac{1}{c}$ in the holographic ( large central charge) limit as described below.

We would like to emphasize here that the exact relation between the holographic negativity and the holographic mutual information described by eq.(\ref {mutual inf}) is valid only for this specific mixed state configuration of adjacent intervals and is not expected to hold for generic mixed states. However it could be shown in \cite {Chaturvedi:2016rcn,Chaturvedi:2016opa} that for a bipartite mixed state involving a single interval the holographic entanglement negativity was described 
by a sum of specific holographic mutual informations between subsystems relevant to the purification. Hence there seems to be
a relation between these two measures in the holographic limit whose specific nature depends on the mixed state configuration in question. A quantum information theoretic understanding of this phenomena is an open issue which needs elucidation.

In the $AdS_3/CFT_2$ scenario being considered here, the bulk dual of the $CFT_{1+1}$ at zero temperature is described by the $AdS_3$ vacuum, whose metric is given as follows
\begin{equation}\label{metric AdS3}
 ds^2 = -\left(\frac{r^2}{R^2}\right)dt^2 +\left(\frac{r^2}{R^2}\right)^ {-1} dr^2 +\left( \frac{r^2}{R^2}\right) d\phi^2,
\end{equation}
where $R$ is the radius of the $\mathrm{AdS_3}$ space time. Employing the conjecture described above \cite {Jain:2017aqk}  the 
holographic entanglement negativity of the configuration in Fig. (\ref{fig2}) may now be obtained as
\begin{equation}\label{53}
\mathcal{E} = \frac{3R}{8G^{(3)}_N} \ln \Big(\frac{l_1l_2}{(l_1+l_2)a}\Big).
\end{equation}
Remarkably the holographic entanglement negativity exactly reproduces the $CFT_{1+1}$ result given in eq. (\ref{cft negativity}) in the large central charge limit \cite{Kulaxizi:2014nma,Jain:2017aqk} upon using the Brown-Henneaux formula ${c}=\frac{3R}{2G^{(3)}_N}$ \cite{Brown:1986nw}.

Note that here we have utilized the relation $r_0\sim \frac{1}{a}$ from the $AdS/CFT$ dictionary that connects the UV cut-off for the boundary $CFT_{1+1}$ to the bulk infra red cut-off ($r_0$) ( to regulate the lengths of geodesics  in eq.(\ref{HEN CONJ1}) ). The eq.(\ref{53}) suggests that only the leading universal part of the negativity in eq.(\ref{cft negativity}) is captured by our conjecture whereas the non-universal constant term is sub leading in the large central charge limit. However the precise renormalization procedure for this is an open issue as the first term in eq.(\ref{cft negativity}) depends on the UV cut-off whereas the non-universal part is a constant. We mention here that the same issue also occurs in the Ryu-Takayanagi conjecture for the holographic entanglement entropy of a single interval in a $CFT_{1+1}$  where the non-universal part is once again a constant (see also \cite{Casini:2006hu,Liu:2012eea,Taylor:2016aoi} for related discussions on renormalized entanglement entropy in higher dimensions). In contrast for higher point twist correlators in a $CFT_{1+1}$ relevant to both the entanglement entropy and the entanglement negativity for multiple intervals involve non universal functions ( of the cross ratios). In this case using monodromy techniques it was clearly demonstrated that the universal parts which admit a bulk geometrical description, are dominant in the large central charge limit \cite{Hartman:2013mia,Malvimat:2017yaj} whereas the non universal functions are sub leading in $\frac {1}{c}$.

\subsection{Entanglement negativity for two adjacent intervals at finite temperature}
For the finite temperature case the entanglement negativity for the mixed state described by the configuration in Fig. (\ref{fig2}) in the context of $CFT_{1+1}$ may be obtained from eq. (\ref{trace 3-point}) through the conformal map $z \rightarrow w = \frac{\beta}{2\pi}~\ln z$ to the cylinder of circumference $\beta$. 
This leads to the following expression for the entanglement negativity 
\begin{equation}\label{negativity fin temp}
\mathcal{E} =
\frac{c}{4} \ln \Bigg(\frac{\beta}{\pi a}\frac{\sinh(\frac{\pi l_1}{\beta})\sinh(\frac{\pi l_2}{\beta})}
{(\sinh\frac{\pi (l_1+l_2)}{\beta})}\Bigg)+ \mathrm{const}, 
\end{equation}
where $\beta=1/T$ and $a$ are the inverse temperature and the UV cut-off in the boundary field theory respectively. As earlier  in the large central charge limit the 'const' term in the above equation may be neglected \cite{Kulaxizi:2014nma,Jain:2017aqk}. 

The bulk dual for the above case is described by the $(2+1)$ dimensional Euclidean BTZ black hole whose metric is
\begin{equation}\label{BTZ metric}
ds^2 = \frac{\left(r^2 - r_{+}^2\right)}{R^2}d\tau^2 +\frac{R^2}{\left(r^2 - r_{+}^2\right)}dr^2 + \frac{r^2}{R^2}d\phi^2.
\end{equation}
Here the horizon radius $r_+$ is related to the inverse Hawking temperature as $\beta=2\pi R^2/r_+$. The holographic entanglement negativity for the two adjacent intervals at a finite temperature may then be obtained from the conjecture eq. (\ref{HEN CONJ1}) proposed in \cite {Jain:2017aqk}. Interestingly as earlier
this exactly reproduces the finite temperature $CFT_{1+1}$ result given by eq. (\ref{negativity fin temp})  in the large central charge limit upon using the Brown-Henneaux formula.
\section{Holographic entanglement negativity for $\mathrm{AdS_{d+1}/CFT_d}$}

In this section we establish the holographic entanglement negativity conjecture for a mixed state described by two adjacent subsystems in a boundary $CFT_d$ in the context of the $AdS_{d+1}/CFT_d$ scenario which was alluded to in the article \cite {Jain:2017aqk}. As mentioned in the Introduction this would involve an algebraic sum of the areas of the bulk co-dimension two extremal surfaces anchored on the respective subsystems. From the conjecture described in \cite {Jain:2017aqk} the holographic entanglement negativity may then be expressed as follows
\begin{equation}\label{HEN CONJ AREA}
\mathcal{E} = \frac{3}{16G^{(d+1)}_N}\big(\mathcal{A}_{1}+\mathcal{A}_{2}-\mathcal{A}_{12}\big),  
\end{equation}
where $\mathcal {A}_i$ is the extremal area of the co-dimension two surface anchored on the subsystem $A_i$. Using the Ryu-Takayanagi prescription \cite{Ryu:2006ef,Ryu:2006bv}, it is possible to express the holographic entanglement negativity in the following form
\begin{equation}\label{mutual inf1}
\mathcal{E} =  \frac{3}{4}(S_{A_1}+S_{A_2}-S_{A_1\cup A_2})=\frac{3}{4}[{\cal I}(A_1,A_2)],
\end{equation}
where $S_{A_i}$ is the holographic entanglement entropy of the subsystem $A_i$. Interestingly the expression in eq. (\ref {mutual inf1}) is the holographic mutual information between the two adjacent subsystems modulo a constant factor. We are now ready to employ our holographic conjecture to evaluate the entanglement negativity for two adjacent subsystems described by $(d-1)$-dimensional spatial rectangular strip geometries in the boundary $CFT_d$ which we will describe in the subsequent sections.

\section{Holographic entanglement negativity for $\mathrm{AdS_{d+1}/CFT_d}$ in vacuum}

\begin{figure}[H]
\centering
\includegraphics[scale=.30 ]{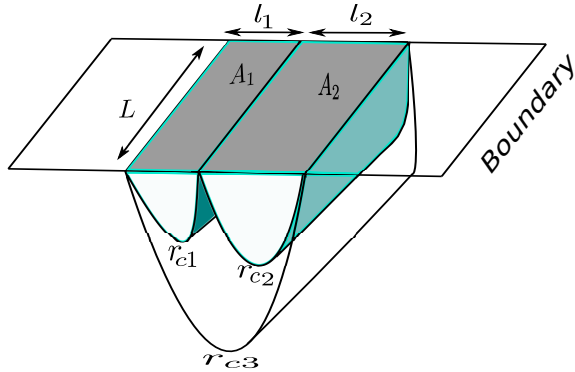}
\caption{Schematic of the extremal surfaces that are anchored on the subsystems $A_1$, $A_2$ and $A_1 \cup A_2 $ involved in the computation of the holographic entanglement negativity of the adjacent subsystems in the boundary $\mathrm{CFT}_d$.}\label{figure1}	
\end{figure}

As mentioned above in this section we now proceed to the computation of the holographic entanglement negativity for two adjacent subsystems described by rectangular strip geometries in the boundary $CFT_d$ at zero temperature. The corresponding bulk dual geometry in this case is the $AdS_{d+1}$ vacuum space time
whose metric in the Poincare coordinates is given as 
\begin{equation}
ds^2=\frac{1}{z^2}\Big(-dt^2+\sum_{i=1}^{d-1}dx_i^2+dz^2\Big),
\end{equation}
where the $AdS$ radius has been set to $R=1$. The respective rectangular strip geometries 
of the two subsystems $A_1$ and $A_2$ depicted in Fig. (\ref{figure1}) are then specified as follows
\begin{equation}
x=x^1 \equiv [-\frac{l_j}{2},\frac{l_j}{2}]~~x^i=[-\frac{L}{2},\frac{L}{2}],~~i=2,...,(d-1),~~j=1,2.
\end{equation}

We now briefly describe the computation for the areas of bulk co-dimension two extremal surfaces anchored on 
rectangular strip geometries in the boundary $CFT_d$ \cite {Ryu:2006ef}. The corresponding 
area functional is expressed in the following way
\begin{equation}
{\cal A}= L^{d-2}\int_{-l/2}^{l/2} dx \frac{\sqrt{1+(\frac{dz}{dx})^2}}{z^{d-1}}.
\end{equation}
The Euler-Lagrange equation for the extremization problem is then given as

\begin{equation}\label{zst}
\frac{dz}{dx}=\frac{\sqrt{z_{*}^{2(d-1)}-z^{2(d-1)}}}{z^{d-1}},
\end{equation}
where $z=z_*$ is the turning point of the extremal surface. The extremal area may then be described as
\begin{equation}\label{area}
\mathcal{A}=\frac{2}{d-2}\Big(\frac{L}{a}\Big)^{d-2}- 2 I \Big(\frac{L}{z_*}\Big)^{d-2},
\end{equation}
where $a$ is the UV cutoff and the constant $I$ is given as
\begin{equation}\label{I}
I=\frac{1}{d-2}-\int_0^1\frac{dy}{y^{d-1}}\Big(\frac{1}{\sqrt{1-y^{2(d-1)}}}-1\Big)=-\frac{\sqrt{\pi}~\Gamma\Big(\frac{2-d}{2(d-1)}\Big)}{\Gamma\Big(\frac{1}{2(d-1)}\Big)}.
\end{equation}
Using the eq.(\ref{zst}), eq. (\ref{area}) and eq. (\ref{I}) it is now possible to express the area of the extremal co-dimension two surface as
\begin{equation}
{\cal A}={\cal A}_{div}-s_0~\Big(\frac{L}{l}\Big)^{d-2},
\end{equation}
where the divergent part of the area ${\cal A}_{div}$ and the constant $s_0$ are given as 
\begin{equation}\label{s0}
\begin{aligned}
s_0&=\frac{2^{d-1}\pi^{(d-1)/2}}{d-2}\Bigg(\frac{\Gamma(\frac{d}{2(d-1)})}{\Gamma(\frac{1}{2(d-1)})}\Bigg)^{d-1},\\
{\cal A}_{div}&=\frac{2}{d-2}\Big(\frac{L}{a}\Big)^{d-2}.
\end{aligned}
\end{equation}
One may now determine the holographic entanglement negativity $\mathcal{E}$ for the mixed state at zero temperature in the boundary $CFT_d$ described by the two strip geometries in Fig. (\ref{figure1}) to be as follows 
\begin{equation}\label{henvacuum}
\begin{aligned}
\mathcal{E}=&\frac{3}{16 G_N^{(d+1)}}\Bigg[\frac{2}{d-2}\Big(\frac{L}{a}\Big)^{d-2}-s_0\bigg\{\Big(\frac{L}{l_1}\Big)^{d-2}+\Big(\frac{L}{l_2}\Big)^{d-2}-\Big(\frac{L}{l_1+l_2}\Big)^{d-2}\bigg\}\Bigg].
\end{aligned}
\end{equation}
The first term in the above expression is the divergent term which is proportional to the area of the entangling surface between the two spatial strips on the $d$ dimensional boundary and the second term
describes the finite part of the negativity.
\section{Holographic entanglement negativity for $\mathrm{AdS_{d+1}/CFT_d}$ at finite temperature}
At finite temperatures the boundary $CFT_d$ is dual to the $AdS_{d+1}$-Schwarzschild black hole with the following metric where the $AdS$-radius has been set to $R=1$
\begin{equation}
  ds^2= -r^2\Big(1-\frac{r_h^d}{r^d}\Big)dt^2+\frac{dr^2}{r^2\Big(1-\frac{r_h^d}{r^d}\Big)}+r^2d\vec{x}^2.
 \end{equation}
The horizon radius $r_h$ is related to the Hawking temperature as $T=r_h d/4\pi$ and $ \vec{x}\equiv(x,x^i) $ are the coordinates on the boundary. We first briefly review the computation for the area ${\cal A}$ of the bulk $AdS_{d+1}$ co-dimension two extremal surface anchored on a single rectangular strip on the boundary as described in \cite{Fischler:2012ca}. This will be subsequently employed to compute the holographic entanglement negativity for the configuration Fig. (\ref{figure1}) in question.
The extremal area functional anchored on a single rectangular strip is given as
\begin{equation}
 {\cal A}=L^{d-2}\int dr r^{d-2}\sqrt{r^2x'^2+\frac{1}{r^2(1-\frac{r_h^d}{r^d})}}.\label{gena}
\end{equation}
The corresponding Euler-Lagrange equation for the extremization problem leads to the following
\begin{equation}\label{tur}
 \frac{l}{2}=\frac{1}{ r_{c} } \int^1_{0} \frac{u^{d-1} du}{\sqrt{(1-u^{2d-2})} }(1-\frac{r_h^d}{r_{c}^d}u^d)^{-\frac{1}{2}},~~~u=\frac{r_c}{r},
\end{equation}
where $r_c$ as earlier describes the turning point. The area functional in terms of the variable $u$ may now be expressed as follows

\begin{equation}\label{ar}
{\cal A} = 2 L^{d-2}r_{c}^{d-2} \int^1_{0} \frac{du}{u^{d-1}\sqrt{(1-u^{2d-2})} }(1-\frac{r_h^d}{r_{c}^d}u^d)^{-\frac{1}{2}}.
\end{equation}
This leads us to the final expression for the area functional as

\begin{equation}\label{fullarea}
{\cal A}=\big(\mathcal{A}_{div}+\mathcal{A}_{finite}\big),
\end{equation}
where $\mathcal{A}_{div}$ is the temperature independent divergent part and $\mathcal{A}_{finite}$ is the finite part. These may be expressed as follows
\begin{align}
\label{afinite}
\begin{split}
\mathcal{A}_{div}&=\frac{2}{d-2}\Big(\frac{L}{a}\Big)^{d-2},\\
\mathcal{A}_{finite}&=2 L^{d-2}r_c^{d-2}\Bigg[\frac{\sqrt{\pi}\Gamma\Big(-\frac{d-2}{2(d-1)}\Big)}{2(d-1)\Gamma\Big(\frac{1}{2(d-1)}\Big)}+\sum_{n=1}^{\infty}\Big(\frac{1}{2(d-1)}\Big)\frac{\Gamma\Big(n+\frac{1}{2}\Big)\Gamma\Big(\frac{d(n-1)+2}{2(d-1)}\Big)}{\Gamma\big(1+n\big)\Gamma\Big(\frac{dn+1}{2(d-1)}\Big)} \Big(\frac{r_h}{r_c}\Big)^{nd} \Bigg].
\end{split}
\end{align}
Note that $r_c>r_h$ from \cite{Hubeny:2012ry} ensuring the convergence of the series in $\mathcal{A}_{finite}$.
The holographic entanglement negativity for the mixed state described by the two intervals in the boundary $CFT_d$ ( Fig. (\ref{figure1}) ) may then be obtained from our conjecture eq. (\ref{HEN CONJ AREA}) as
\begin{equation}
\begin{split}
{\cal E} &= \frac{3}{16G_{N}^{(d+1)}} \Bigg[\frac{2}{d-2}(\frac{L}{a})^{d-2}\\
&+2 L^{d-2}r_{c1}^{d-2}\bigg\{\frac{\sqrt{\pi}\Gamma\Big(-\frac{d-2}{2(d-1)}\Big)}{2(d-1)\Gamma\Big(\frac{1}{2(d-1)}\Big)}+\sum_{n=1}^{\infty}\Big(\frac{1}{2(d-1)}\Big)\frac{\Gamma\Big(n+\frac{1}{2}\Big)\Gamma\Big(\frac{d(n-1)+2}{2(d-1)}\Big)}{\Gamma\big(1+n\big)\Gamma\Big(\frac{dn+1}{2(d-1)}\Big)} \Big(\frac{r_h}{r_{c1}}\Big)^{nd} \bigg\}\\
&+2 L^{d-2}r_{c2}^{d-2}\bigg\{\frac{\sqrt{\pi}\Gamma\Big(-\frac{d-2}{2(d-1)}\Big)}{2(d-1)\Gamma\Big(\frac{1}{2(d-1)}\Big)}+\sum_{n=1}^{\infty}\Big(\frac{1}{2(d-1)}\Big)\frac{\Gamma\Big(n+\frac{1}{2}\Big)\Gamma\Big(\frac{d(n-1)+2}{2(d-1)}\Big)}{\Gamma\big(1+n\big)\Gamma\Big(\frac{dn+1}{2(d-1)}\Big)} \Big(\frac{r_h}{r_{c2}}\Big)^{nd} \bigg\}\\
&-2L^{d-2}r_{c3}^{d-2}\bigg\{\frac{\sqrt{\pi}\Gamma\Big(-\frac{d-2}{2(d-1)}\Big)}{2(d-1)\Gamma\Big(\frac{1}{2(d-1)}\Big)}+\sum_{n=1}^{\infty}\Big(\frac{1}{2(d-1)}\Big)\frac{\Gamma\Big(n+\frac{1}{2}\Big)\Gamma\Big(\frac{d(n-1)+2}{2(d-1)}\Big)}{\Gamma\big(1+n\big)\Gamma\Big(\frac{dn+1}{2(d-1)}\Big)} \Big(\frac{r_h}{r_{c3}}\Big)^{nd} \bigg\}\Bigg].
 \end{split}
\end{equation}
Here $r_{c1},r_{c2},r_{c3}$ are the turning points of the extremal surfaces in the bulk  anchored on the strips $A_1,A_2$ and $A_1\cup A_2$ on the boundary respectively. It is required to evaluate the quantity $r_{ci}$ from the eq. (\ref{tur}) in terms of $l_i$ and $r_h$. The corresponding integral is not analytically solvable but may be determined perturbatively for low and high temperature approximations described in the following subsections.
\subsection{Holographic entanglement negativity in the low temperature limit}
At low temperature we have $Tl\ll1~(r_hl\ll1)$ and $r_c$ may be determined perturbatively as an expansion in $r_hl$ \cite{Fischler:2012ca}, which leads to the finite part of the area (\ref{afinite}) as
\begin{equation}\label{afinite2}
\mathcal{A}_{finite}=s_0\Big(\frac{L}{l}\Big)^{d-2}\bigg[1+s_1 (r_hl)^d+O[ (r_hl)^{2d} ]\bigg].
\end{equation}
Here the constants $s_0$ is given by the eq. (\ref{s0}) and $s_1$ is given as
\begin{equation}\label{afinite1}
s_1=\frac{\Gamma\big(\frac{1}{2(d-1)}\big)^{d+1}}{2^{d+1}\pi^{\frac{d}{2}}\Gamma\big(\frac{d}{2(d-1)}\big)^d\Gamma\big(\frac{d+1}{2(d-1)}\big)}\bigg(\frac{\Gamma\big(\frac{1}{d-1}\big)}{\Gamma\big(-\frac{d-2}{2(d-1)}\big)}+\frac{2^\frac{1}{d-1}(d-2)\Gamma\big(1+\frac{1}{2(d-1)}\big)}{\sqrt{\pi}(d+1)}\bigg).
\end{equation}

\begin{figure}[H]
\centering
\includegraphics[scale=.3]{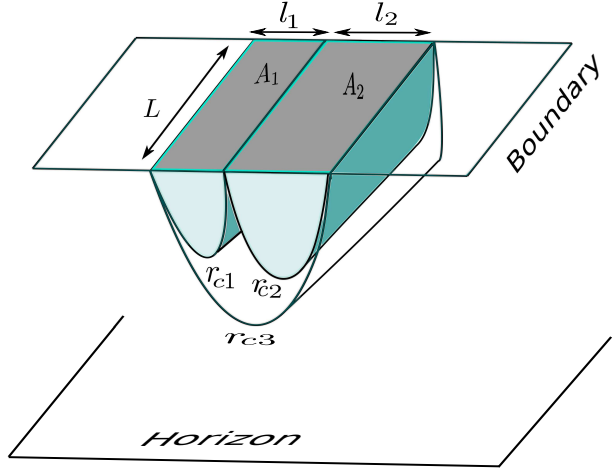}
\caption{Schematic of the extremal surfaces that are anchored on the subsystems $A_1$, $A_2$ and $A_1 \cup A_2 $ in the boundary $\mathrm{CFT}_d$ at low temperatures.}\label{figure2}	
\end{figure}

The holographic entanglement negativity at low temperature for the mixed state in the boundary $CFT_d$ described by the configuration in Fig (\ref{figure2}) may now be obtained from our conjecture (\ref{HEN CONJ AREA}) and the Eqns (\ref{afinite2}) and (\ref{afinite1}) in the following way
\begin{eqnarray}
\mathcal{E}=\frac{3}{16 G_N^{(d+1)}}\Bigg[&&\frac{2}{d-2}\Big(\frac{L}{a}\Big)^{d-2}+s_0\Bigg\{\Big(\frac{L}{l_1}\Big)^{d-2}+\Big(\frac{L}{l_2}\Big)^{d-2}-\Big(\frac{L}{l_1+l_2}\Big)^{d-2}\Bigg\}\nonumber\\&&-k~l_1 l_2L^{d-2}T^d+s_0\Bigg\{\Big(\frac{L}{l_1}\Big){\cal O}\big(Tl_1\big)^{2d}+\Big(\frac{L}{l_2}\Big){\cal O}\big(Tl_2\big)^{2d}\Bigg\}\Bigg].~~~~~~\label{hen at inite temp}   
\end{eqnarray}
Here the constant $k$ is given as
\begin{equation}
k=2 (\frac{4 \pi}{d})^d~ s_0s_1.
\end{equation}
Note that the first and the second term in the above expression are temperature independent describing the contribution to the holographic entanglement negativity from the $AdS$ vacuum eq. (\ref{henvacuum}). The remaining terms are the finite temperature corrections to the holographic entanglement negativity at low temperatures for the boundary $CFT_d$.
\subsection{Holographic entanglement negativity in the high temperature limit}
For high temperatures we have $Tl\gg1~(r_hl\gg1)$ and in this case it is possible to obtain the quantity $r_c$ eq. (\ref{tur}) in a near horizon expansion in $\epsilon=(\frac{r_c}{r_h} -1)$ \cite{Fischler:2012ca} as follows

\begin{equation}\label{rc}
r_c=r_h(1+\epsilon).
\end{equation}
Here $\epsilon$ is expressed as 
\begin{align}\label{eent}
\begin{split}
\epsilon=&C_1\exp(-\sqrt{\frac{d(d-1)}{2}}l r_h),\\
\end{split}
\end{align}
where the constant $C_1$ is given as
\begin{align}\label{eent1}
\begin{split}
C_1=&\frac{1}{d}\ \exp\Bigg[\sqrt{\frac{d(d-1)}{2}}\Bigg\{\frac{2\sqrt{\pi}\Gamma\Big(\frac{d}{2(d-1)}\Big)}{\Gamma\Big(\frac{1}{2(d-1)}\Big)}\\
&+2\sum_{n=1}^{\infty} \Bigg(\frac{1}{1+nd}\frac{\Gamma\Big(\frac{1}{2}+n\Big)\Gamma\Big(\frac{d(n+1)}{2(d-1)}\Big)}{\Gamma\Big(1+n\Big)\Gamma\Big(\frac{dn+1}{2(d-1)}\Big)}-\frac{1}{\sqrt{2d(d-1)}~n}\Bigg)\Bigg\} \Bigg].
\end{split}
\end{align}

The area of the extremal surface at high temperatures is expressed as
\begin{align}\label{ee}
\begin{split}
\mathcal{A}=&\frac{2}{d-2}\Big(\frac{L}{a}\Big)^{d-2}+\Big(\frac{4\pi}{d}\Big)^{d-1}\Bigg[V~T^{d-1}+\frac{C_2 ~d}{8\pi}A'~T^{d-2}\\
&-\frac{C_1}{8\pi}\sqrt{2d(d-1)}~A'~T^{d-2} ~exp~\Big\{-\sqrt{(d-1)/2d}~4 \pi Tl\Big\}+...\Bigg],
\end{split}
\end{align}
where $V=l~L^{d-2}$ and $A'=2L^{d-2}$ are the volume and area of a single strip respectively . The constant term $C_2$ is given as
\begin{align}
\begin{split}
C_2&=2\Bigg[-\frac{\sqrt{\pi}(d-1)\Gamma\Big(\frac{d}{2(d-1)}\Big)}{(d-2)\Gamma\Big(\frac{1}{2(d-1)}\Big)}+\sum_{n=1}^{\infty}\frac{1}{1+n d}\Big(\frac{d-1}{d(n-1)+2}\Big)\frac{\Gamma\Big(n+1/2\Big)\Gamma\Big(\frac{d(n+1)}{2d-2}\Big)}{\Gamma\big(n+1\big)\Gamma\Big(\frac{dn+1}{2d-2}\Big)}\Bigg].
\end{split}
\end{align}

\begin{figure}[H]
\centering
\includegraphics[scale=.3 ]{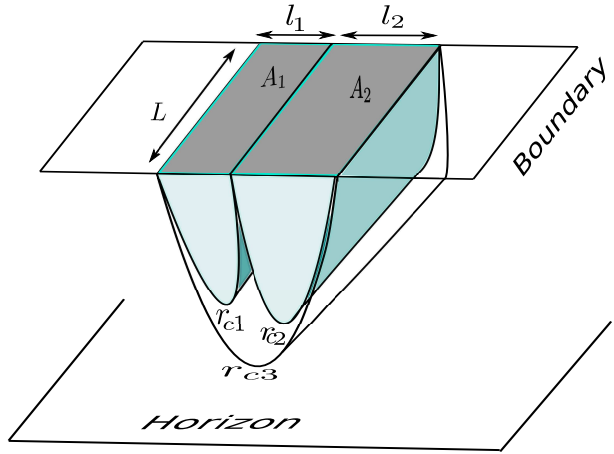}
\caption{Schematic of the extremal surfaces that are anchored on the subsystems $A_1$, $A_2$ and $A_1 \cup A_2 $ in the boundary $\mathrm{CFT}_d$ at high temperatures.}\label{figure3}	
\end{figure}

The holographic entanglement negativity at high temperatures for the mixed state in the boundary $CFT_d$ described by the configuration in Fig. (\ref{figure3}) may then be established from the eq. (\ref{ee}) employing our conjecture as follows
\begin{align}\label{enht}
\begin{split}
{\cal E}=&\frac{3}{16 G_N^{(d+1)}}\frac{2}{(d-2)}\Big(\frac{A}{a^{d-2}}\Big)+\frac{3}{16 G_N^{(d+1)}}\Big(\frac{4\pi}{d}\Big)^{d-1}\Bigg[\frac{C_2~d}{4\pi}A~T^{d-2}\\
&-\frac{C_1}{4\pi}\sqrt{2d(d-1)}~A~T^{d-2}~\Bigg\{ exp~\Big(-\sqrt{(d-1)/2d}~4 \pi Tl_1\Big)+exp~\Big(-\sqrt{(d-1)/2d}~4 \pi Tl_2\Big)\\
&-exp~\Big(-\sqrt{(d-1)/2d}~4 \pi T(l_1+l_2)\Big)\Bigg\}+...\Bigg],
\end{split}
\end{align}
where the ellipsis represent the higher order corrections and $A=L^{d-2}$ is the area of the entangling surface shared by the two adjacent strips on the boundary. Interestingly in the above expression notice that 
the thermal contribution to the holographic entanglement negativity (proportional to the volume in the eq. (\ref{ee})) has been subtracted out rendering it to be proportional to the area of the entangling surface. This is in conformity with the usual expectations from quantum information theory and furthermore, recently it has been demonstrated that entanglement negativity does obey an area law in various many body systems such as the finite temperature quantum spin model and the two dimensional harmonic lattice in \cite{DeNobili:2016nmj,PhysRevE.93.022128}.
\section{Summary and conclusions}

To summarize we have established a holographic conjecture for the entanglement negativity for mixed states of adjacent subsystems in zero and finite temperature
boundary $CFT_d$s. The relevant subsystems are described by $(d-1)$-dimensional spatial rectangular strip geometries at the boundary in a $AdS_{d+1}/CFT_d$ scenario. Our conjecture involves a certain algebraic sum of the areas of bulk co dimension two extremal surfaces anchored on the corresponding subsystems on the $AdS_{d+1}$ boundary and was motivated by the corresponding analysis for the $AdS_3/CFT_2$ scenario in \cite {Jain:2017aqk}. It is interesting that the algebraic sum described above actually characterizes the holographic mutual information between the two adjacent subsystems. Note that these two  measures are completely distinct quantities in quantum information theory. Our conjecture states that only their universal parts (which are dominant in the holographic limit) match upto a numerical factor for the particular mixed state configuration of adjacent subsystems. We emphasize that such a matching between the universal parts of negativity and mutual information of adjacent intervals in a $CFT_{1+1}$ has also been demonstrated for both local and global quench problems in \cite{Coser:2014gsa,Wen:2015qwa}.

The holographic entanglement negativity for the boundary $CFT_d$ at zero temperature could then be computed from the bulk dual geometry described by the $AdS_{d+1}$ vacuum from our conjecture. The corresponding holographic entanglement negativity for the boundary $CFT_d$ at finite temperature however involved a bulk dual geometry described by the $AdS_{d+1}$-Schwarzschild black hole with a planar horizon. In the latter case the area integrals are not analytically solvable and were evaluated in a perturbative expansion for low and high temperatures. It was observed from our computation that the leading contribution to the holographic entanglement negativity at low temperature arises from the $AdS_{d+1}$ vacuum with subleading thermal corrections. Interestingly on the other hand at high temperatures the finite part of the holographic entanglement negativity is proportional to the area of the entangling surface on the boundary whereas the volume dependent thermal parts cancel out. It has been demonstrated that entanglement negativity does obey such area laws in various condensed matter systems  confirming the expectation from quantum information theory (See \cite{DeNobili:2016nmj,PhysRevE.93.022128}).   

Through these examples we demonstrated that our conjecture provides a direct and elegant holographic prescription to compute the entanglement negativity for mixed states described by the specific configuration in boundary $CFT_d$ both at zero and finite temperatures. However for the higher dimensional $AdS_{d+1}/CFT_d$ scenario this remains a conjecture. So in higher dimensions our conjecture requires further analysis towards a possible proof from the bulk side which remains a non trivial open issue. In this context our examples serve as a first consistency check in higher dimensions and lead to interesting results described above. 

It is well known from quantum information theory that the entanglement negativity characterizes the upper bound on the distillable entanglement for mixed states. It is expected that our conjecture will lead to a deeper understanding of entanglement issues for diverse applications in higher dimensional conformal field theories from condensed matter physics to quantum gravity. It would be interesting to compute the holographic entanglement negativity for subsystems described by more general geometries other than the rectangular strip geometries considered by us. This would possibly lead to deeper insights into the nature of holographic quantum entanglement and its relation to issues of quantum gravity. We expect to return to these exciting issues in the near future.

\section{Acknowledgment}
Parul Jain would like to thank Prof. Mariano Cadoni for his guidance and 
the Department of Physics, Indian Institute of Technology Kanpur, India for their warm hospitality. Parul Jain's work is financially supported by Universit\`a di Cagliari, Italy and INFN, Sezione di Cagliari, Italy.
%
\bibliographystyle{utphys}

\bibliography{HENSchAdS}

\end{document}